# F-802.11p: A Fuzzy Enhancement for IEEE 802.11p in Vehicle-to-Everything Communications

Hamdy A.M. Sayedahmed[1], Emadeldin M. Elgamal[2,1] and Hesham A. Hefny[1]

[1]Department of Computer Science,
Faculty of Graduate Studies for Statistical Research, Cairo University, Egypt
[2]Department of Engineering and Computer Science,
Tarleton State University, Texas, Stephenville, USA

## ABSTRACT

*Vehicle-to-Everything communications (V2X) are becoming increasingly popular as a solution for safer roads and better traffic management. One of the essential protocols in V2X is the Dedicated Short Range Communication (DSRC) protocol suite. DSRC includes the IEEE 802.11p protocol that operates at the medium access control (MAC) and physical (PHY) layers. Upon collision, the IEEE 802.11p MAC layer applies a carrier sense multiple access/collision avoidance (CSMA/CA) mechanism that randomly selects a backoff time to re-check the channel activity and then retransmit. However, the random selection of the backoff time may lead to further packet collisions that decrease the utilization of the communication channel, which suffers from a limited bandwidth in the first place. This paper proposes a fuzzy model based on rational decision-making, which we call F-802.11p, to improve the IEEE 802.11p protocol backoff time selection by limiting the IEEE 802.11p beacon messages to better use of the available bandwidth. A simulation study presents the evaluation of our work compared to IEEE 802.11p. We deployed the simulation software in two scenarios: the Veins Framework map and the map of New Administrative Cairo in Egypt. We base our comparison on slots backoff, times into backoff, PHY busy time, MAC busy time, total lost packets, and generated/received beacon messages. Simulation results show that both protocols have comparable results in slots backoff, times into back off, and the generated beacon messages. At the same time, our F-802.11p significantly outperforms the IEEE 802.11p in PHY busy time, MAC busy time, total lost packets, and the received beacon messages in both scenarios.*

## 1. INTRODUCTION

Vehicle-to-Everything communications (V2X) can be supported by either the 3GPP LTE, cellular V2X (C-V2X), or the IEEE 802.11p access layer. Based on the IEEE 802.11p, two standards are used to support V2X: the dedicated short-range communication (DSRC) developed by the U.S. and the Intelligent Transportation System (ITS)-G5 developed by the European Telecommunications Standards Institute (ETSI). Our focus in this work is on DSRC. DSRC is designed to provide vehicle-to-vehicle (V2V) and vehicle-to-infrastructure (V2I) direct communications [22]. DSRC employs Wireless Access in Vehicular Environments (WAVE) that contains the IEEE 1609 family, which includes IEEE 1609.4 for multichannel operations, and IEEE 802.11p, which operates at the media access control (MAC) and physical (PHY) layers. IEEE 1609.4 divides the allocated spectrum into six service channels (SCH) and one control channel (CCH). Furthermore, it divides channel time access into 50 *ms* CCH interval and 50 *ms* SCH interval as sequential synchronous intervals [8, 12]. On the other hand, IEEE 802.11p MAC depends on the enhanced distributed channel access (EDCA) mechanism that allows vehicles to access the channel based on a Distribution Coordination Function (DCF). In this scheme, a vehicle senses the channel before transmitting, and if the channel is idle for the arbitration interframe space (AIFS) period, it accesses the channel directly. Otherwise, a carrier sense multiple access/collision avoidance (CSMA/CA) MAC protocol is enabled in which a backoff time is randomly selected from the interval [0, W], where W is the current vehicle contention window (C.W.). Initially, C.W. is set to its minimum value ($CW_{min}$) and doubles on sequent collisions until a maximum contention window $CW_{max}$ is reached. The selection of the $CW_{min}$ and $CW_{max}$ values depends on the application type, which includes best-effort traffic (B.E.), background traffic (B.K.), video traffic (VI), and voice traffic (V.O.) [33].

Communications in DSRC encompass many message types that include regular DSRC packets, WAVE Short messages (WSMs), Basic Safety Messages (BSMs), and WAVE Service Advertisement messages (WSAs). WSMs, also known as beacons, are broadcast messages sent periodically at a rate of 2-10 Hz by a vehicle to its neighbors to inform the current state. BSMs are used to communicate special and dangerous traffic situations such as intersection collisions or roadside alerts. WSAs convey information about services such as traffic alerts, tolling, navigation, restaurant information, entertainment, and Internet access. The number of WSM, BSM, and WSA messages is directly proportional to vehicle density. If a channel is busy, collisions are potentially arising. Consequently, in a high dense network and limited bandwidth, message reliability and latency constraints may not be reached [1, 5, 9, 11] [14 - 15] [17 - 18] [37].

Typically, vehicle density exhibits uneven distribution and is dynamic. Vehicles may be condensed in a small area, causing high interference, or may form a sparse network in that each vehicle is out of range of the others. The dynamic nature of vehicles' movements makes it difficult to determine their precise speeds, signal strengths, and distances in real-time, complicating building accurate topology control platforms [2-3]. IEEE 802.11p MAC applies CSMA/CA, where the C.W. selection does not adapt to V2X dynamicity. As a result, a vehicle may wait for a long time due to a busy channel state. Even worse, a vehicle in a transmitted mode may have collided packets, doubling its C.W. During these long waiting times, a vehicle cannot send BSM to declare an accident and cannot send warning messages.

This paper aims to enhance the IEEE 802.11p MAC and PHY layers by exploiting the uncertainty in V2X to self-adapt to the different conditions to achieve the reliability and delay requirements for various V2X applications. We propose a fuzzy model, which we call F-802.11p, to decide on the backoff time instead of selecting it randomly as implemented in the IEEE 802.11p. The reason behind using fuzzy logic has discussed in section 3. To evaluate our proposal, we conducted simulation studies that compare F-802.11p against the IEEE 802.11p in two scenarios: the Veins Framework map and the map of New Administrative Cairo in Egypt. Results from the simulation show that both protocols have comparable results in slots backoff, times into back off, and the generated beacon messages, while F-802.11p significantly outperforms the IEEE 802.11p in PHY busy time, MAC busy time, total lost packets, and the received beacon messages. The main contribution of this work is summarized as follows:

- Proposing a fuzzy model (F-802.11p) to enhance the MAC and PHY layers to reduce generated/received beacon packets in most network conditions to decrease collisions.
- Performing a simulation-based validation using Simulation of Urban Mobility (SUMO)0.30.0, Veins 5.0, Objective Modular Network Testbed in C++ (OMNeT++) 5.4.1, Open Street Map, and MATLAB 2016b fuzzy toolbox.

- Analyzing the effect of the fuzzy model on backoff parameters (C.W. and AIFS) in DSRCstandards based on the simulation results.

The rest of the paper is organized as follows. A review of related work is given in Section 2. Section 3 introduces our F-802.11p proposal. The simulation environment is presented in Section 4. Section 5 discusses the simulation results. Section 6 is the conclusion.

## 2. RELATED RESEARCH

Many researchers addressed the calculation of the contention window on collision, and several articles have been introduced to replace the IEEE 802.11p random selection of the backoff time. Objectives of earlier work in this area fall in the following directions:

- Analyze the IEEE 802.11p MAC layer to gain insights into packet delivery ratio and backoff duration [13, 33].
- Enhancing the protocol performance through improving random contention window selection [20, 22], channel access mechanisms [38], throughput [25], collision probabilities [23], and congestion [16].
- Optimizing the channel access [27] [21].
- Ensuring the delivery of safety messages [7].
- Fairness of the time-slot allocation [35] and solving the unbalanced backoffs times [10].

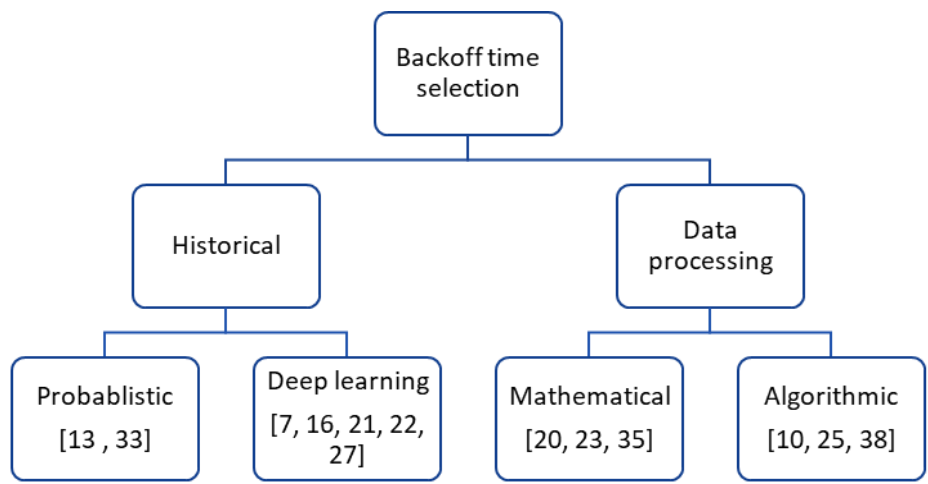

Figure 1. Classification of Back off time selection techniques.

Such earlier efforts can be broadly classified into historical-based or data processing based. On the one hand, historical data techniques, in a broad context, are based on collecting data about past events and circumstances of a particular subject. On the other hand, historical-based methods can be further divided into probabilistic and deep learning models. Probabilistic techniques use statistics to examine data. They assume randomness to predict future events based on past ones. Deep learning (DL), referred to as features representation learning is the relationship between features using hierarchal representation. Based on artificial machine learning (ML), DL and neural networks (ANNs) are provided with features and some known training examples.

On the other hand, data processing schemes rely on a real-time collection of data about events or actions. Data processing-based techniques can be classified into mathematical and algorithmic. Mathematical models describe a real-world problem in mathematical terms, typically in the form of equations. They use these equations to help understand the original problem and discover new features about the situation. An algorithm carefully crafts a set of instructions that take specific inputs to achieve a particular task. Figure 1 presents a classification of the backoff time selection algorithms.

To evaluate their work, earlier efforts used simulators and numerical analysis. Simulations that were used include such ns-2 [7, 25, 35], ns-3 [1, 10, 20, 21], iTETRIS [16], OPNET [23], OMNeT++ [27] along with SUMO, Veins and OpenStreetMap. Numerical analysis is basically performed based on MATLAB [13, 38] and the probabilistic PRISM model [33].

Several works adopted fuzzy logic to enhance IEEE 802.11p by relieving contention window random size [40–41, 44-49, 51], controlling beacon messages [39, 50, 52-53], and managing the MAC layer [42-43]. However, to the best of the authors' knowledge, no previous work evaluated the effect of a fuzzy model in IEEE 802.11p with V2V and V2I environments considering contention window random size, controlling beacon messages, managing MAC layer, and PHY engaged time simultaneously. Table 1 shows a feature-based comparison between our proposed model with other related works that adopted fuzzy logic.

Table 1. Comparison with other works adopting fuzzy logic. C.W.: contention windows random size issue, B.M.: controlling beacon messages.

| Ref. | Network type | CW | BM | MAC layer | PHY Layer |
|---|---|---|---|---|---|
| [40, 41, 44, 45, 46, 47, 48, 49, 51] | V2V | √ | | | |
| [39, 50, 52, 53] | V2V | | √ | | |
| [42, 43 ] | V2V & V2I | | | √ | |
| Our proposed model | V2V & V2I | √ | √ | √ | √ |

## 3. METHODOLOGY AND PROPOSED MODEL

This paper proposes F-802.11p, a fuzzy enhancement for the IEEE 802.11p. The IEEE 802.11p requires senders to wait on packet collision for a contention window (C.W.) that is selected at random. The random selection of the C.W. may not be the best strategy as a small C.W. may lead to further collisions, and a large C.W. may lead to unnecessary long waiting times. However, several parameters such as speed, signal-to-noise ratio, sender gain, receiver gain, power level, the distance between vehicles and RSU, and channel idle time are studied. We conducted the assessment study of these parameters using a k-means clustering algorithm. Our assessment study determined the most influential parameters like speed, sender gain, and receiver gain. Furthermore, our F-802.11p increases the vehicle's channel idle time by exploiting the uncertainty of input parameters to reduce the number of collided and redundant packets and network overhead. More specifically, F-802.11p utilizes the vehicle's speed, sender's gain, and receiver's gain inputs to enable vehicles to decide rationally in sending response/request WSM, WSA, and BSM beacon messages.

With the F-802.11p, and before broadcasting a packet, a vehicle checks its status in terms of three parameters: speed, gain, and the nearest vehicle gain (receiver) to optimize channel idle time. Typically, these parameters change at a rate of milliseconds. The vehicle's speed changes because of road speed limits, frequent stops due to traffic lights, and environmental hazards. The sending and receiving vehicle gain are hard to measure because of obstacles and signal interference. If a vehicle's status matches the fuzzy model linguistic terms, it can broadcast its message. Otherwise, the vehicle waits. The fuzzy model creates an arrangement pattern of sending and receiving packets, and it allows vehicles with the same status to communicate reliably and prevents the vehicles with the undesired status from engaging the channel with redundant packets.

Due to the uncertainty in the three parameters, we use fuzzy logic as a method for representing and processing the uncertainty based on "degrees of truth" rather than conventional computing with a discrete outcome of "true or false" or "1 or 0" to provide low-cost approximate solutions to relieve C.W. randomization. Generally, a fuzzy model deals with imprecise data based on expert knowledge to help control unpredictable systems to control accuracy. Here, the proposed model deals with a lack of exact knowledge (speed, sender gain, receiver gain) to reach an approximate reasoning decision to achieve better vehicle channel consumption.

Each transmitted message contains a set of attributes such as sender gain, receiver gain, and speed. Grouping, feature selection, forming, and cleaning these attributes' values to make a dataset have been made according to [6]. The dataset shows a relationship between the vehicle's speed, sender gain, receiver gain, and channel idle time. Also, the fuzzy model knowledge base (IF…THEN…rules) is based on this dataset. The knowledge base is based on expert knowledge that a domain expert or fuzzy clustering could obtain. The collected dataset is clustered by fuzzy clustering to extract the knowledge base in this work.

Our performance metrics include the following:

- Slots backoff and times into backoff, which are measures of how many times a vehicle invoked C.W.
- Physical layer busy time and MAC busy time represent each vehicle channel busy time.
- Total lost packets and generated/received beacon messages (WSM, BSM, WSA) are measures of how many times a vehicle can send/receive packets on its channel with no packet repetition.

For performance evaluation, we compare our model against the standard IEEE 802.11p in two different maps. The first map is the default Veins' framework map, and the second is the New Administrative capital city in Egypt. In both maps, we follow [22, 27, 31, 33] to configure the number of RSU, interval, and length of a beacon, data length, and mobility model. Simulation of Urban Mobility (SUMO) 0.30.0 is used to simulate road traffic, and Objective Modular Network Testbed in C++ (OMNeT++) 5.4.1 is used to simulate V2V and V2I networks; Veins 5.0 is a framework for vehicular networks that is based on OMNeT++ and SUMO. To import the second map, we used Open Street Map (OSM), a simulator for a geographical database that is based on Google maps. Finally, we used MATLAB 2016b fuzzy toolbox to create the fuzzy model.

Figure 2 shows the fuzzy model processing when creating or receiving a packet. First, a vehicle treats the upper message by checking that the message is not scheduled and is not a self-message. Then, if the message type is an upper beacon message, the vehicle senses the channel by sending a request to other vehicles.

### 3.1. Fuzzy Model F-802.11p

A fuzzy model is a mathematical representation of a relationship that deals with multi-criteria in vague conditions and contains inputs (I), outputs (O), linguistic terms (representation of labels and intervals), and membership functions (representation of truth degree), and IF…THEN rule-base (representation of knowledgebase) [54]. Herein, the F-802.11p contains three inputs: speed ($s$), sender gain ($S_G$), and receiver gain ($R_G$), one output: channel idle time, referred to as factor (F), and nine IF…THEN…rules. All parameters are normalized in the range [0,1]. The fuzzification membership functions (MFs) used for I and O are triangular and trapezoidal. Equations (1) and (2) show the MFs for triangular and trapezoidal, respectively. Multiple tries give the selection of MF [36]. Also, the linguistic term is accepted when the input is in the range of $TH_i$ and $TH_j$ for triangular or trapezoidal membership function. Table 2 illustrates the proposed model parameters.

$$\mu(m) = \begin{cases} 0 & if\ x \leq TH1 \\ \frac{(x-TH1)}{(Th2-TH1)} & if\ TH1 < x < TH2 \\ 1 & if\ x \geq TH2 \end{cases} \quad (1)$$

Where TH1 is a threshold for the active system, and TH2 is a threshold to identify the level of activeness.

$$\mu(m) = \begin{cases} 0 & if\ x \leq TH1 \\ \frac{(x-TH1)}{(Th2-TH1)} & if\ TH1 < x < TH2 \\ 1 & if\ TH2 < x < TH3 \\ \frac{TH3-x}{TH3-TH2} & if\ TH3 < x < TH4 \end{cases} \quad (2)$$

Where TH1 and TH4 are lower and upper limits, TH2 and TH3 are lower support and upper support limits.

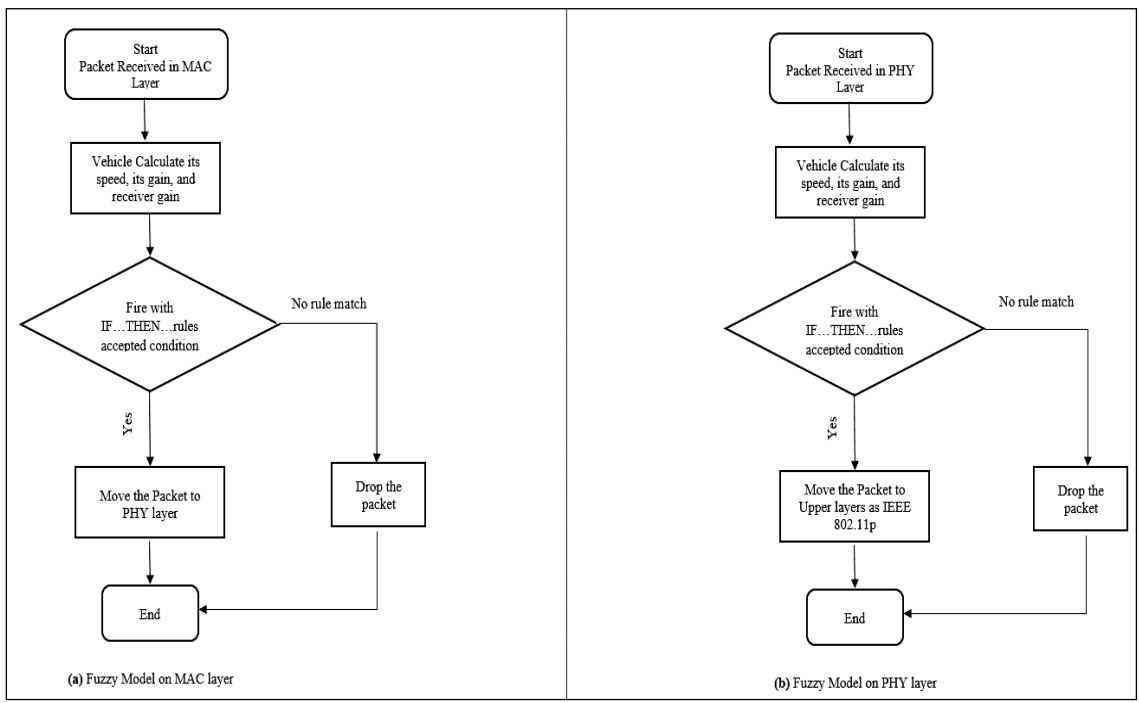

Figure 2. Fuzzy Model in MAC layer (a) and PHY (b) layer.

Table 2. F-802.11p Parameters: R: resident, M: move, N: normal, SL: slow, Fa: fast, W: weak, M: medium, E: Excellent, A: bad, G: good, V.G.: very good

| Parameter Detail | S | | | | | $S_G$ | | | $R_G$ | | | F | | |
|---|---|---|---|---|---|---|---|---|---|---|---|---|---|---|
| Type | I | | | | | I | | | I | | | O | | |
| Linguistic Term | R | M | N | SL | Fa | W | M | E | W | M | E | B | G | V.G. |
| Universe of discourse | 0-8.3 | 5-11.1 | 8.3-19 | 10-22.2 | 13-27.78 | 0-0.4 | 0.1-0.9 | 0.5-1 | 0-0.4 | 0.1-0.9 | 0.5-1 | 0-32.8 | 17-63 | 40-87.1 |
| Membership Function | Tri | Tri | Tri | Tra | Tra | Tri | Tri | Tri | Tri | Tri | Tri | Tri | Tra | Tra |

### 3.1.1. Inputs

The input variables are speed ($S$), sender gain ($S_G$), and receiver gain ($R_G$). The speed is described through five linguistic terms: R = resident, M = move, N = normal, SL = slow, and Fa = fast. Three linguistic terms describe the sender gain and receiver gain: W = weak, M = medium, and E = Excellent. The universe of discourse for S is [0, 27.78], for $S_G$ is [0, 1], and for $R_G$ is [0,1]. The fuzzification membership functions for $S_G$ and $R_G$ are triangular for all linguistic variables. The fuzzification membership functions (MFs) for $S$ are triangular and trapezoidal, as illustrated in Figures 3 and 4.

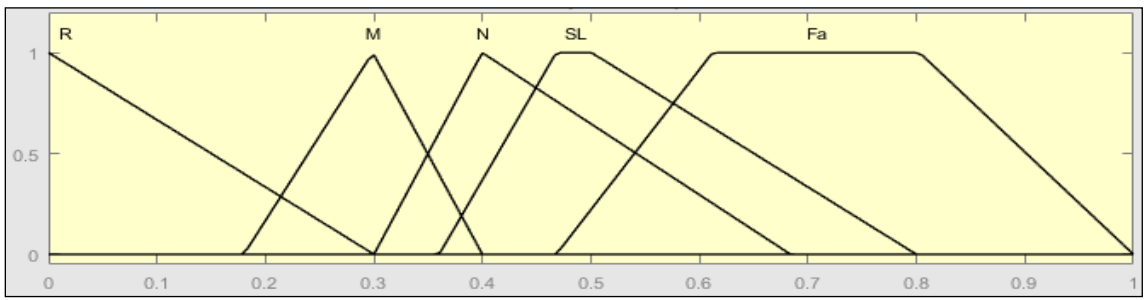

Figure 3. MFs of linguistic terms of Speed (S)

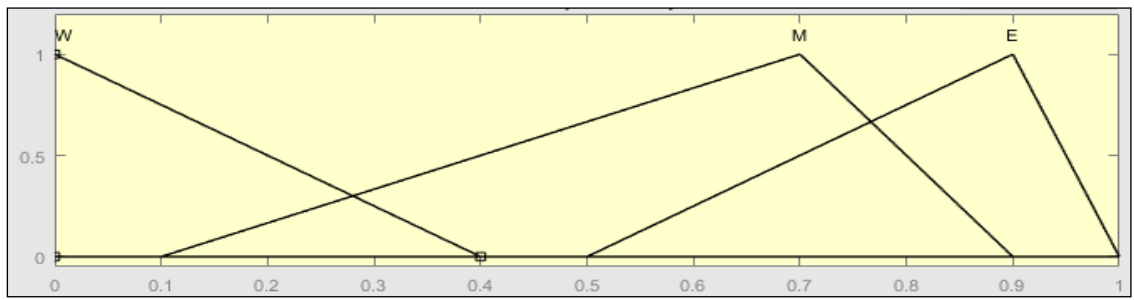

Figure 4. MFs of linguistic terms of sender gain (S.G.) and receiver gain (R.G.)

### 3.1.2. The Fuzzy Rules

The fuzzy IF…THEN…rules represent the knowledge base of the fuzzy model, which could be built by a domain expert or fuzzy clustering [29]. In this work, the presented rules are obtained by the fuzzy C-means algorithm (FCM) [24]. This clustering algorithm drawback is that there is no prior knowledge about the number of clusters. Therefore, an intensive analysis of the number of clusters has been done from cluster numbers 3 to 40. We started with 17 rules and minimized by a distance matrix to nine rules [26, 32]. The fuzzy inference system (FIS) type is Mamdani, the "And" method is min, the "OR" method is max, and the defuzzification is the centroid. Table 3 shows the nine "IF-THEN-Rules".

### 3.1.3. Outputs

The output variable is the factor (F), representing the idle seconds each vehicle could have to transmit beacons. F is described through 3 linguistic terms: A = bad, G = good, VG = V.good. The universe of discourse for F (Channel idle time) is [0, 87.1]. The defuzzification membership functions are trapezoidal for the first and second linguistic term and triangular for the third linguistic term. Figure 5 shows the membership function with linguistic terms.

Table 3. The Nine Fuzzy IF…THEN…Rules

| | | | | |
|---|---|---|---|---|
| R1 | IF Speed is Resident | AND SG is Medium | AND RG is Medium | THEN F is Bad |
| R2 | IF Speed is Move | AND SG is Medium | AND RG is Medium | THEN F is Bad |
| R3 | IF Speed is Normal | AND SG is Medium | AND RG is Medium | THEN F is Bad |
| R4 | IF Speed is Slow | AND SG is Medium | AND RG is Medium | THEN F is Good |
| R5 | IF Speed is Fast | AND SG is Medium | AND RG is Medium | THEN F is Good |
| R6 | IF Speed is Fast | AND SG is Medium | AND RG is Excellent | THEN F is V.Good |
| R7 | IF Speed is Resident | AND SG is Weak | AND RG is Medium | THEN F is Bad |
| R8 | IF Speed is Slow | AND SG is Excellent | AND RG is Weak | THEN F is Bad |
| R9 | IF Speed is Fast | AND SG is Medium | AND RG is Weak | THEN F is Bad |

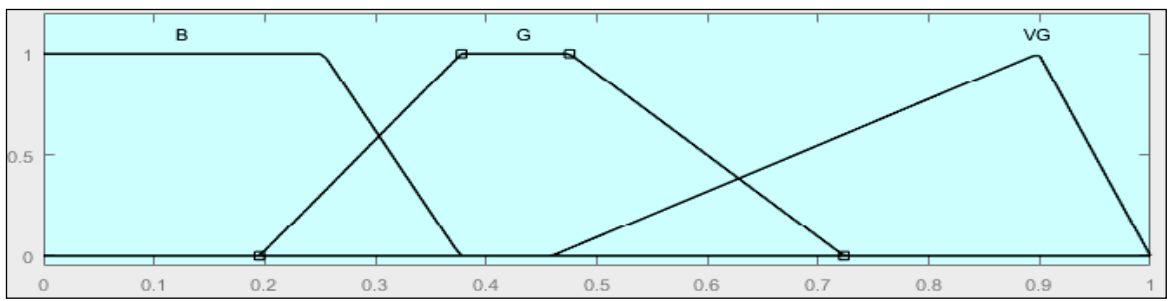

Figure 5. MFs of linguistic terms of the output factor (F)

## 4. EVALUATION ENVIRONMENT

IEEE 802.11p and F-802.11p are evaluated in an environment that includes OMNeT++, SUMO, Veins 5.0 framework for V2X, and MATLAB R2016b fuzzy toolbox in two scenarios. The first scenario uses the Veins framework map with 6 Mbps as a bit rate to compare both protocols and a 20*100 *m*m* area. The vehicles' speed is randomly varied between 0 and 30 km/h with a vehicle's density of 44 that was generated by the framework. For the second scenario simulation area, the existing road network of the New Administrative Capital (NAC) in Egypt is considered. The satellite image of the NAC road network is retrieved from the Open Street Map (OSM). A realistic road traffic scenario is created with the help of Simulation of Urban Mobility (SUMO). The second scenario area is a 100*100 *m*m* area with 27 Mbps as a bit rate, and vehicles' speed was varied between 0 to 80 km/h with a density of 193 that was generated from an actual traffic flow. Also, the second scenario was restricted to the mentioned area due to simulation issues. A simple application has been added to the application layer in both scenarios. Also, F-802.11p is enabled with the acceptance condition "Good" for the first scenario and "V.Good" for the second scenario. Table 4 shows the standard settings of the two scenarios. Both scenarios cover low and high vehicle density and highway and urban areas.

Table 4. Settings of Simulation Environment

| Simulation Parameter | Value |
| --- | --- |
| Simulation time | 200 s |
| Protocol suite | DSRC |
| Protocol | IEEE8021.11p / F-802.11p |
| No. of RSU | 1 |
| Beacon Interval | 1s |
| Beacon Length | 256 bit |
| Data Length | 1024 bit |
| Number of accidents | 10 |
| Contention Windows (min, max) | (15,1023) |
| Backoff slot time | 13 μs |
| Channel frequency | 5.89 GHz |
| Channel bandwidth | 10 MHz |
| Mobility model | Traci model |

## 5. RESULTS AND DISCUSSION

Twelve experiments are performed to compare our proposed F-802.11p against the IEEE 802.11p in the two scenarios presented above. Figures 6 through 17 present the results of the experiments. The discussions of the results of conducted experiments are presented as follows:

**Busy time of MAC and Physical layers**: A vehicle considers a channel idle if the summation of MAC busy time and PHY busy time is idle. The total busy time measures how many seconds the MAC layer treated the channel as busy, while busy time is a PHY layer measure that increases for each frame received above a sensitivity threshold. Figures 6 and 7 show both scenarios' total busy time and busy time. A vehicle drops a packet(s) when it has an out speed normalized range (SL or Fa) and out sender/receiver gain normalized range (M) when using F-802.11p. Thus, the F-802.11p limits the number of processed packets in MAC and PHY layers. However, vehicles in a dense vehicle network satisfy the model constraints as in Figure 6 (b), so they take some time to process the packets. Due to their small numbers, the PHY is processes-free, as illustrated in Figure 7 (b). Therefore, the PHY busy time and the MAC total busy time are reduced consequently, and the F-802.11p outperforms the standard IEEE 802.11p.

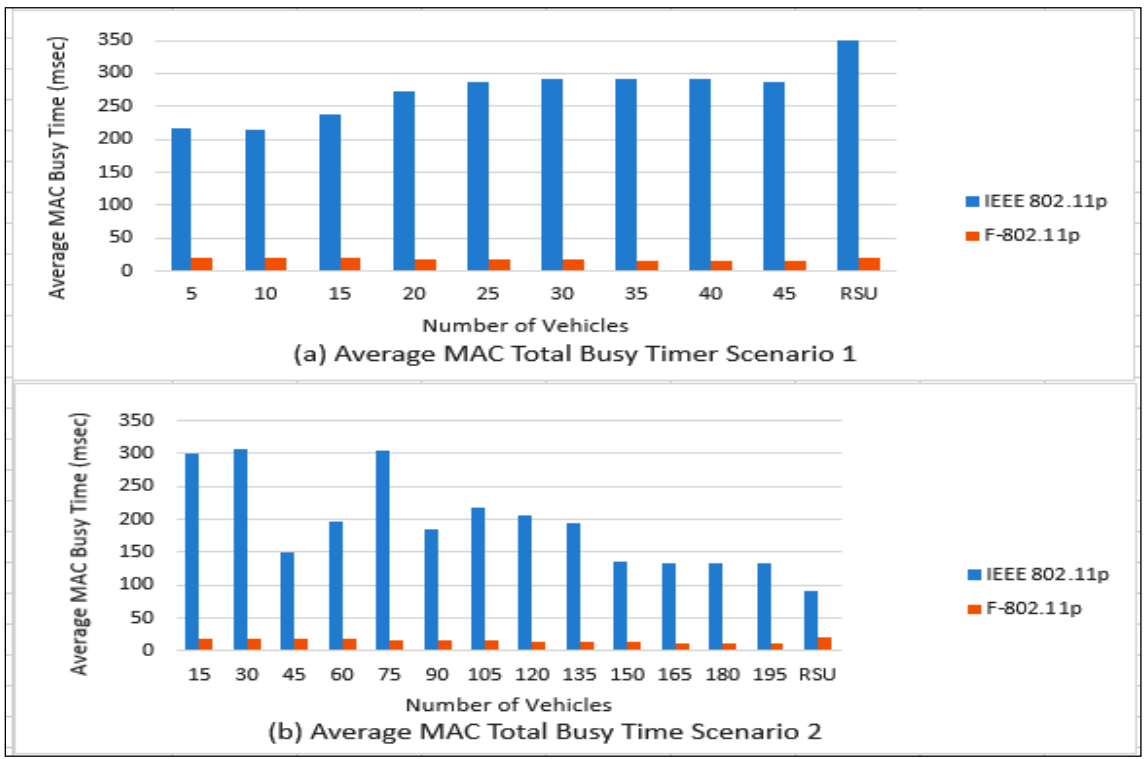

Figure 6. MAC total busy time. (a) Scenario 1, (b) Scenario 2.

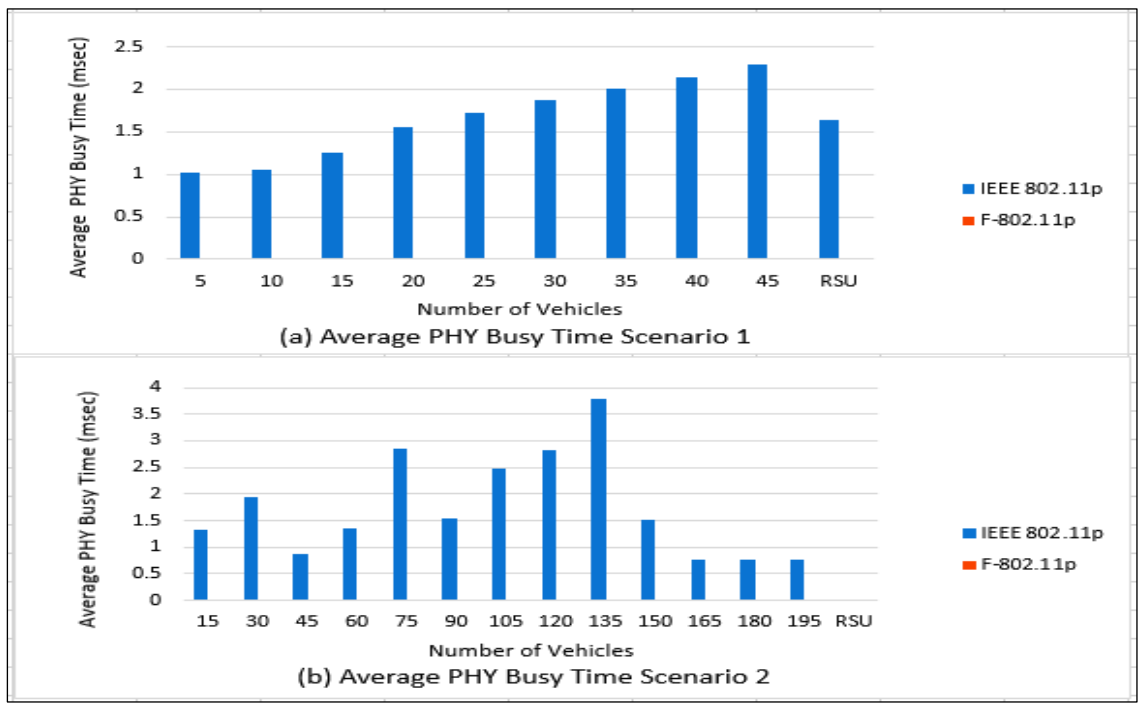

Figure 7. PHY busy time. (a) Scenario 1, (b) Scenario 2

**Generated WSMs, BSMs, and WSAs:** The Wave Short Message (WSM) is a periodic message created by the vehicle to its neighbors with its current state. A basic safety message (BSM) is a non-periodic broadcast message that is triggered by a vehicle to alert dangerous traffic circumstances (intersection collisions). A Wave short advertisement message (WSA) is a non-periodic broadcast message that is created by a vehicle to inform infotainment services. Figures 8, 9, and 10 present the collected results for generated WSMs, BSMs, and WSAs for both scenarios.

Figure 8 (a) shows that the number of WSMs created by IEEE 802.11p is less than those produced by F-802.11p. However, both do not exceed five messages during the simulation interval. Controlling the MAC and PHY layer when transmitting packets drops packets until vehicle(s) is under speed and sender/receiver gain allowed ranges when using F-802.11p requires additional effort to broadcast messages. Therefore, IEEE 802.11p outperforms F-802.11p in a low-dense network. However, RSU creates multiple WSMs in case of using IEEE 802.11p more than those using F-802.11p. Figure 8 (b) contains a dense vehicle network, the performance of F-802.11p is almost similar to IEEE 802.11p except in some vehicles (15, 75, 90). F-802.11p creates no messages due to the organizing of speed and sender/receiver gain allowed range; vehicles have no changes in their states. In contrast, the organizing has a trade-off as in vehicles (60) that requires more messages.

BSM should be delivered to the nearest vehicle(s) of the situation (accident or road collision). Delivering a message to a far or un-participated vehicle consumes bandwidth. Figure 9 (a) shows that the performance of F-802.11p is almost similar to IEEE 802.11p. Figure 9 (b) shows that the number of BSMs created by IEEE 802.11p is more than those produced by F-802.11p because vehicles are under fuzzy model requirements, so they wait to be under speed and sender/gain allowed ranges.

As the number of the requested service(s) by vehicles increases, the number of generated WSAs increases when using F-802.11p. Figure 10 (a) shows that F-802.11p generates an extra message because the vehicle's speed or gain is not in the range of the accepted linguistic term. In Figure 10 (b), vehicle density (60) generates one message more than IEEE 802.11p because some vehicles are not satisfying the fuzzy model accepted conditions. However, vehicles (30, 45, 105, 120, 150) F-802.11p is almost similar to IEEE 802.11p. In contrast, F-802.11p outperforms the standard IEEE 802.11p in vehicles (15, 75, 90, 135, 165, 180, 195) because some vehicles require more messages to get service due they are out of communication range of the requested vehicle or their packets have collided.

**Sent packets**: Sent packets (SP) include packets sent during message dissemination, generated WSMs, WSAs, and BSMs. Figure 11 shows the SP for both scenarios. The SP using F-802.11p is greater than the IEEE 802.11p due to fuzzy model controls vehicle transmitting packets with speed (10-27.78) for the first scenario and speed for the second scenario (13-27.78). However, RSU in Figure 11 (a) and vehicles (165, 180, 195) in Figure 11 (b) that use the IEEE 802.11p need additional packets because of the increasing time for backoff when sensing a channel busy.

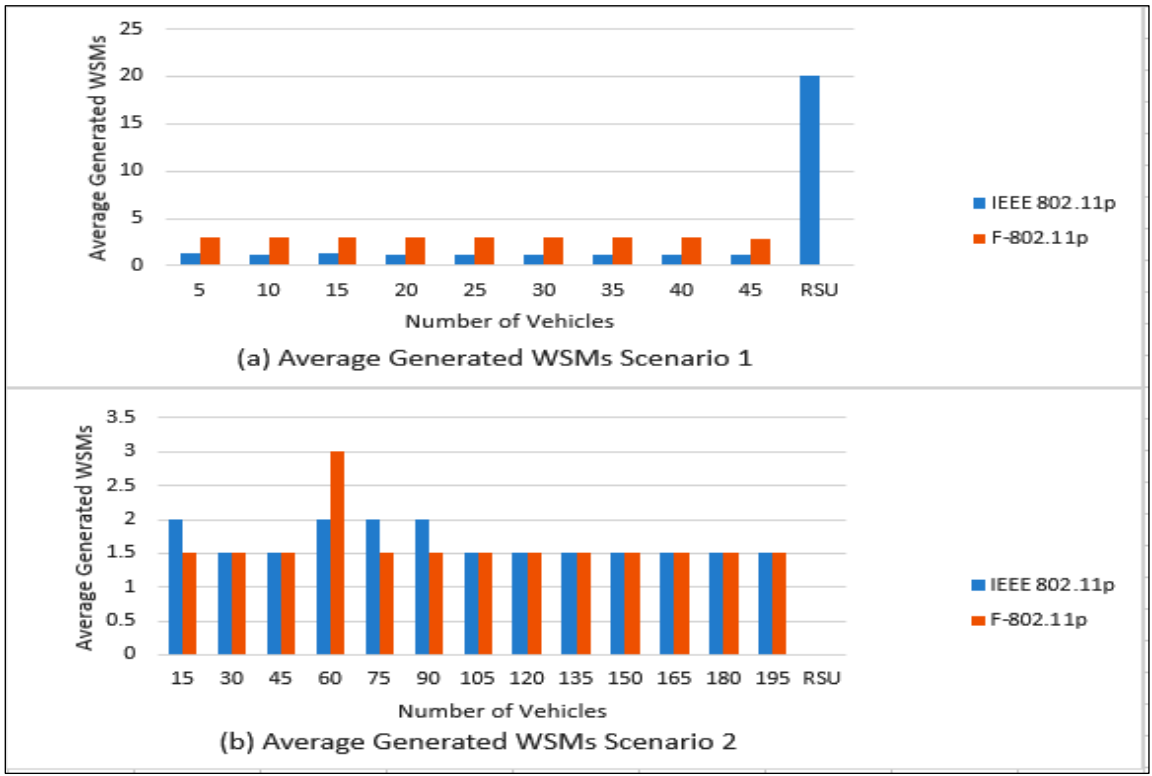

Figure 8. The Generated WSMs. (a) Scenario 1, (b) Scenario 2

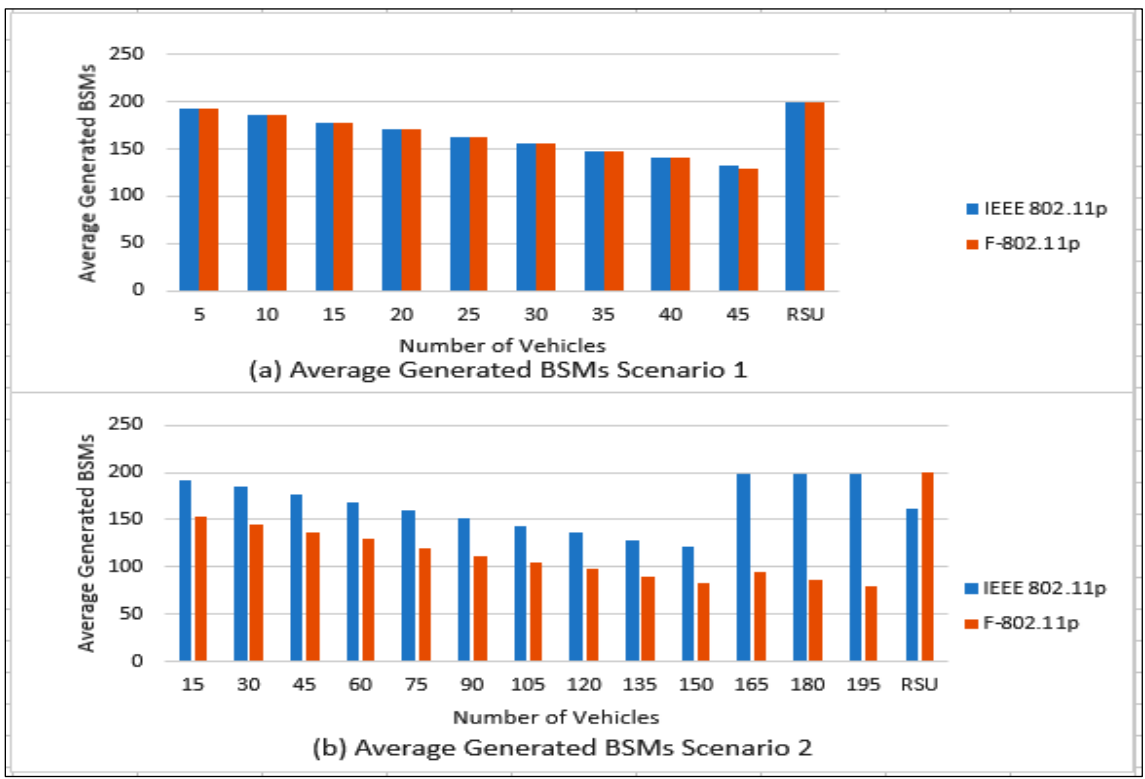

Figure 9. The Generated BSMs. (a) Scenario 1, (b) Scenario 2

International Journal of Computer Networks & Communications (IJCNC) Vol.14, No.4, July 2022

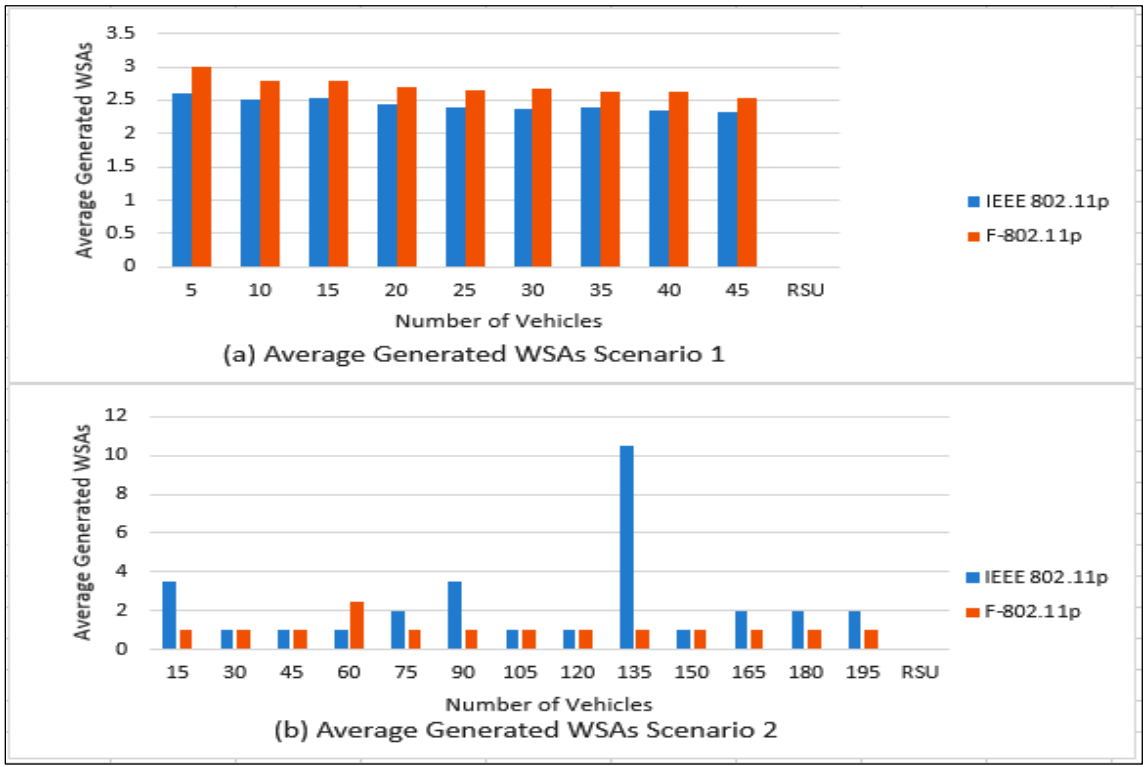

Figure 10. The Generated WSAs. (a)Scenario 1, (b) Scenario 2

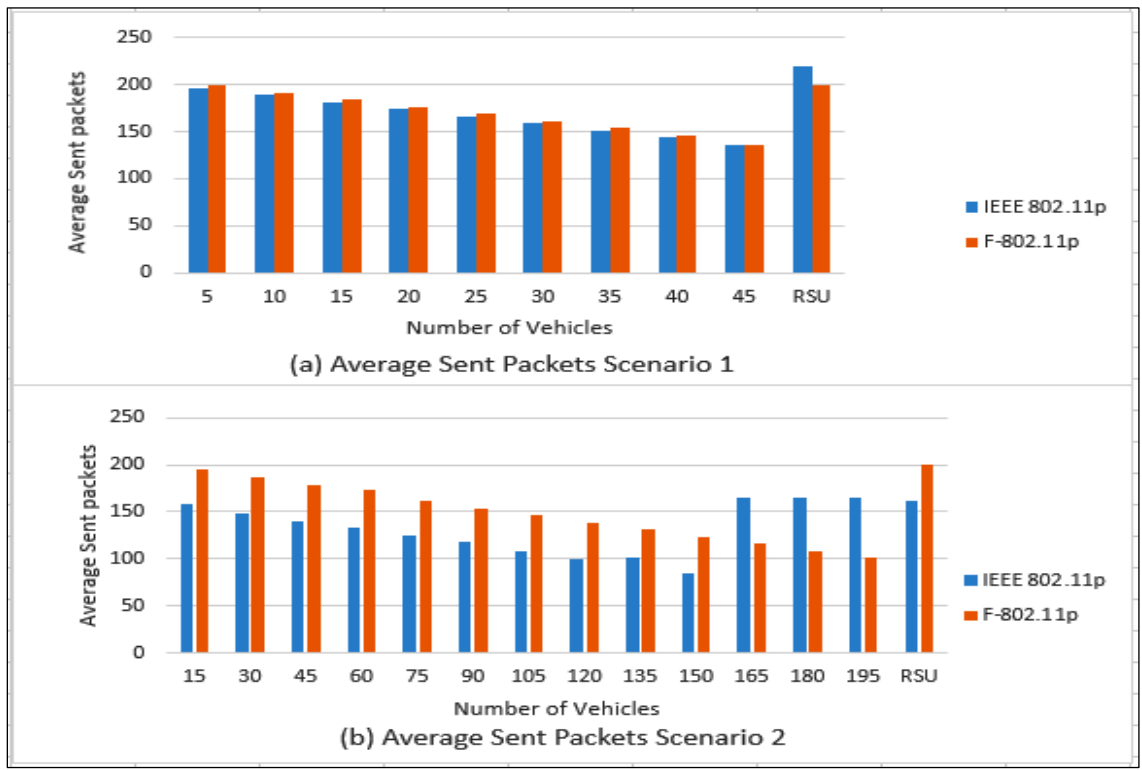

Figure 11. The sent Packets. (a) Scenario 1, (b) Scenario 2

31

**Received WSM, BSM, and WSA**: Received WSM is a metric for how frequently vehicles change their states. Received BSM measures how often vehicles are informed of a critical situation(s). Received WSA presents the summation of informing requests about services. However, these broadcast messages strongly affect backoff and channel activity. Figures 12, 13, and 14 present both scenarios' collected results for received WSMs, BSMs, and WSAs.

In Figure 12 (a), all vehicle's received other vehicle's change states, and due to no acknowledgment messages and broadcasting procedures, vehicles receive duplicate WSMs in the case of IEEE 802.11p. Vehicles that use F-802.11p cannot accept such messages because the model requires all vehicles to be in a range of speed and a range of sender/receiver gain, i.e., the vehicle states kept known with no changes. Consequently, the model drops initial and redundant packets. In Figure 12 (b), the behavior of F-802.11p is as scenario 1 considers sender/receiver gain. When a vehicle finds receiver gain lower than a threshold, it waits and keeps the channel idle. Vehicles that use IEEE 802.11p receive changes on other vehicles' states because the broadcasted vehicle is in the communication range of other vehicles with no threshold.

In Figure 13 (a), all vehicles are informed of dangerous situations from other vehicles, and due to no acknowledgment messages and broadcasting procedures, vehicles receive duplicate BSMs in the case of IEEE 802.11p. Due to the model limitations, vehicles that use F-802.11p cannot receive such messages. If a situation occurs, the broadcasted vehicle is in the same direction as the informed vehicles. Hence, there is no need for broadcasting or in another direction as informed vehicles and also no need for broadcasting. In Figure 13 (b), F-802.11p considers sender/receiver gain. Thus when it finds that the receiver gain is lower than a threshold, it waits and keeps the channel idle. When using IEEE 802.11p, the average received BSMs is reduced because the broadcasted vehicle communication range cannot cover other vehicles.

In Figure 14 (a), all vehicles received WSAs because of the broadcasting procedures. Vehicles receive duplicate WSAs in the case of IEEE 802.11p. Vehicles that use F-802.11p cannot accept such messages because of the model requirements (speed and gain). In Figure 14 (b), F-802.11p considers sender/receiver gain; thus, when it finds receiver gain is lower than a threshold ($TH_i$, $TH_j$), it waits and keeps the channel idle. When using IEEE 802.11p, the average received BSMs is reduced because vehicles are out of the range of the broadcasted vehicle.

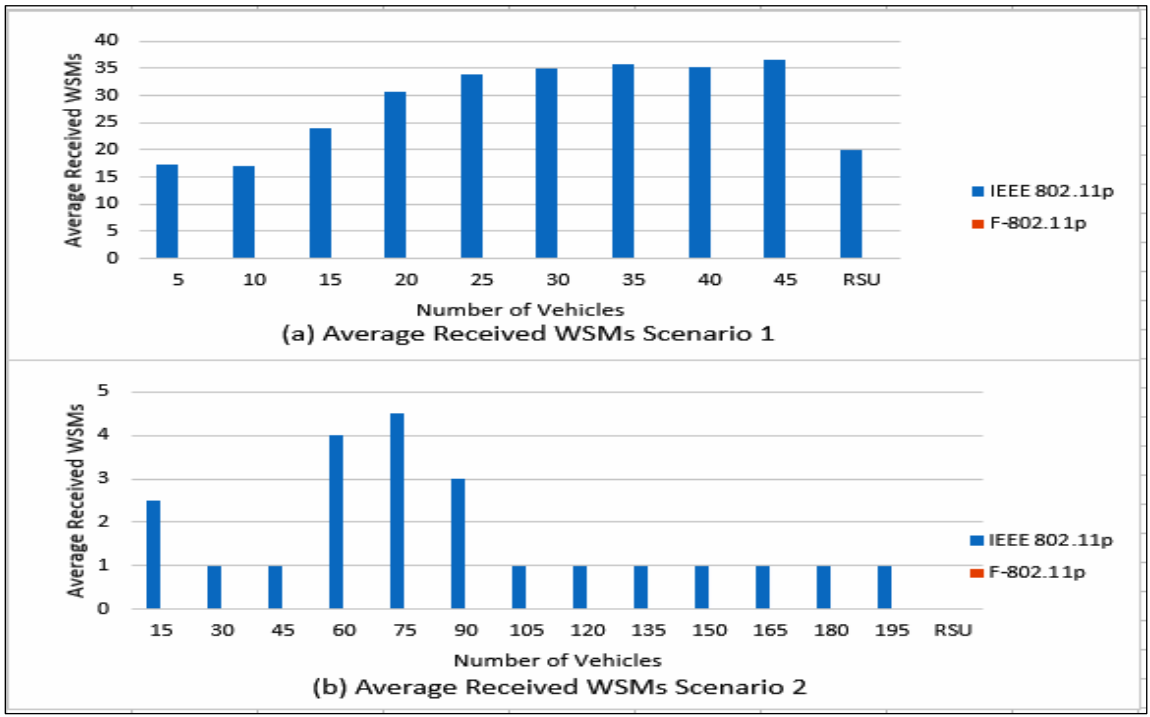

Figure 12. The Received WSMs. (a) Scenario 1, (b) Scenario 2

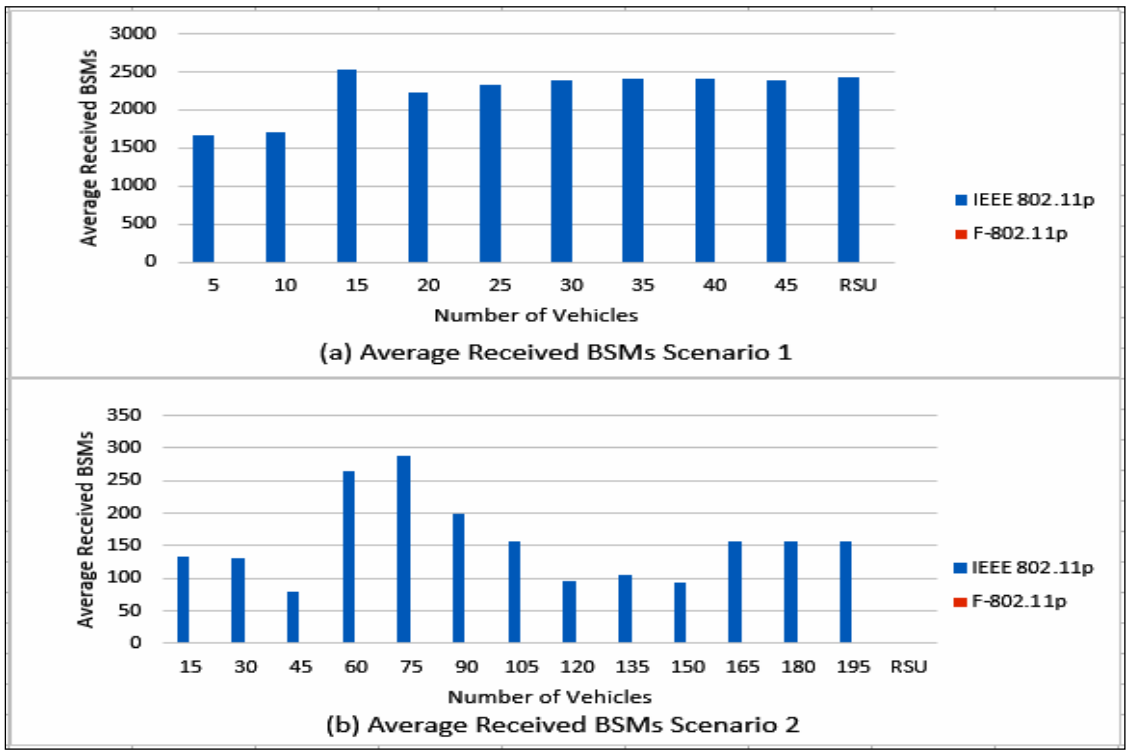

Figure 13. The Received BSMs. (a) Scenario 1, (b) Scenario 2

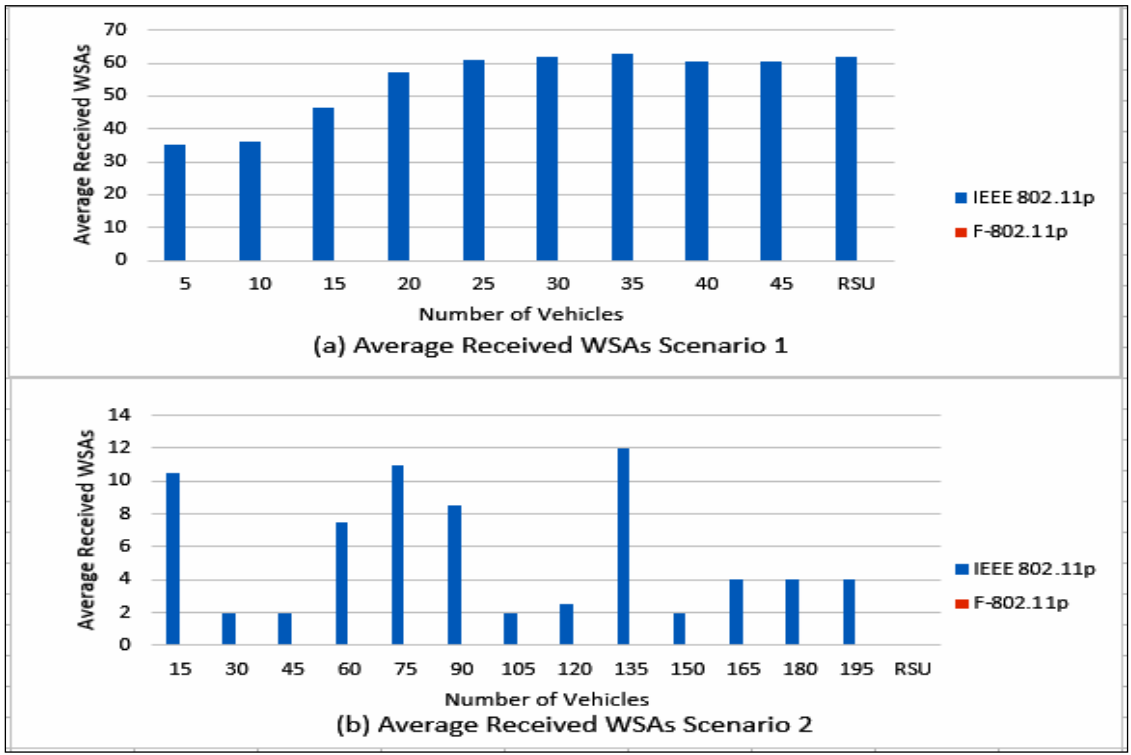

Figure 14. The Received WSAs. (a) Scenario1, (b) Scenario 2

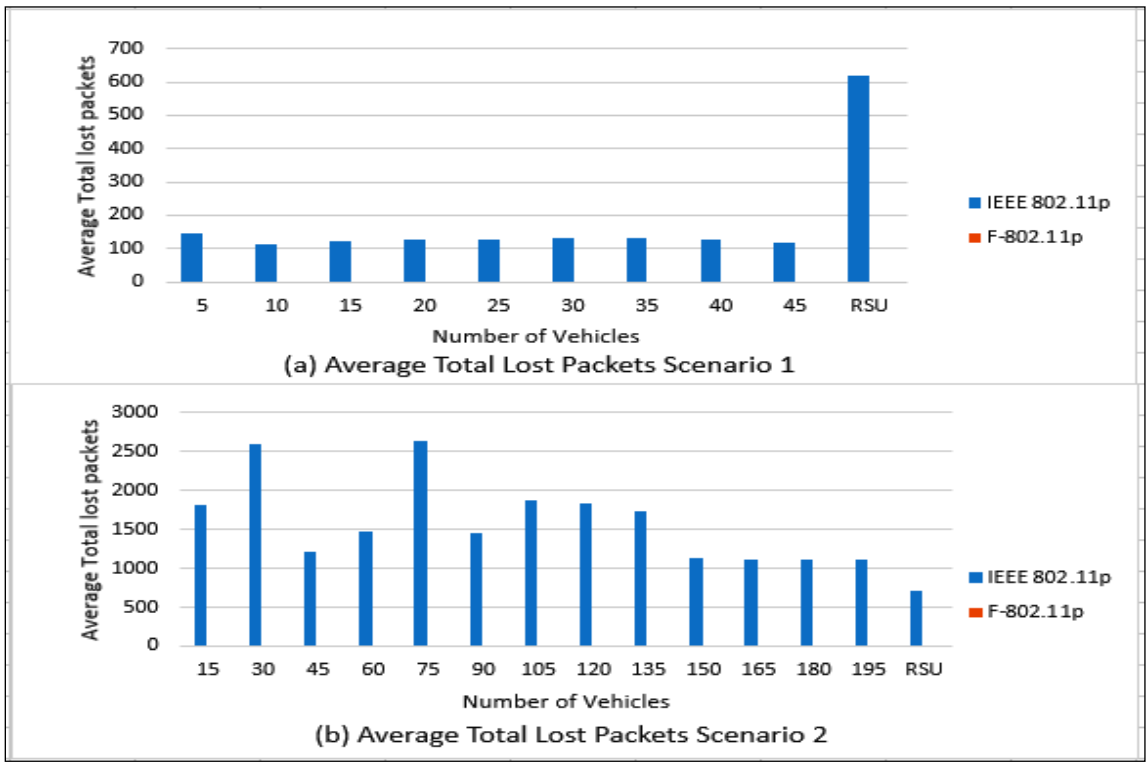

Figure 15. Total Lost Packets. (a) Scenario 1, (b) Scenario 2

34

**Total lost packets**: Total lost packets were the summation of lost packets when the interference occurred, either bit error or collision. Figure 15 depicts the F-802.11p and IEEE 802.11p collected results for both scenarios. Figure 15 (a) and (b) shows that F-802.11p outperforms IEEE 802.11p. Using F-802.11p increases the channel idle time by reducing redundant packets, reducing the number of collided packets, and decreasing the bit error rate. Using IEEE 802.11p, each vehicle receives a broadcast and rebroadcasts the message resulting in more message collisions or interruptions.

**Time into backoff**: Time into backoff is the number of times the vehicle is in backoff. Slots backoff is the number of slots due to backoff. Time into backoff and slots backoff has a strong positive correlation with CSMA/CA MAC protocol as they express how often a channel sensed busy. Figures 16 and 17 show the time into backoff and slots backoff for both scenarios. In Figure 16 (a), and due to F-802.11p limitations, vehicles are not satisfying the conditions of the fuzzy model while they sense a channel idle. Therefore, they require several backoff times more than traditional IEEE 802.11p. Vehicles (15) and RSU get into backoff when using IEEE 802.11p because a channel is sensed busy. Generally, F-802.11p acts as IEEE 802.11p in times into backoff. The behavior of F-802.11p and IEEE 802.11p in Figure 16 (a) is similar to that in Figure 16 (b).

Vehicles may select the same slot backoff size, and in transmitting, their packets will collide in the medium. In Figure 17 (a), most vehicles that use F-802.11p require additional slots than IEEE 802.11p because they are under fuzzy model constraints that need vehicles to wait until they reach accepted condition boundaries. However, RSU using IEEE 802.11p does more slots than the fuzzy model because it frequently senses the channel busy. The behavior of F-802.11p and IEEE 802.11p in Figure 17 (a) is similar to that of Figure 17 (b).

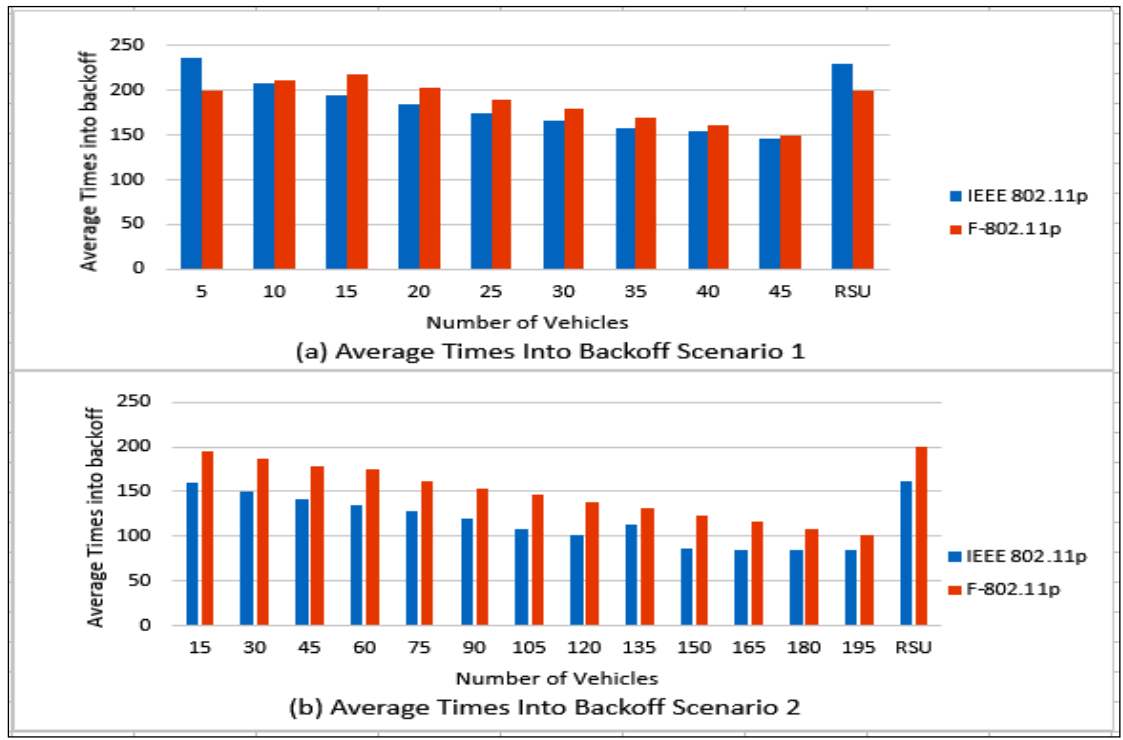

Figure 16. Times into backoff. (a) Scenario 1, (b) Scenario 2

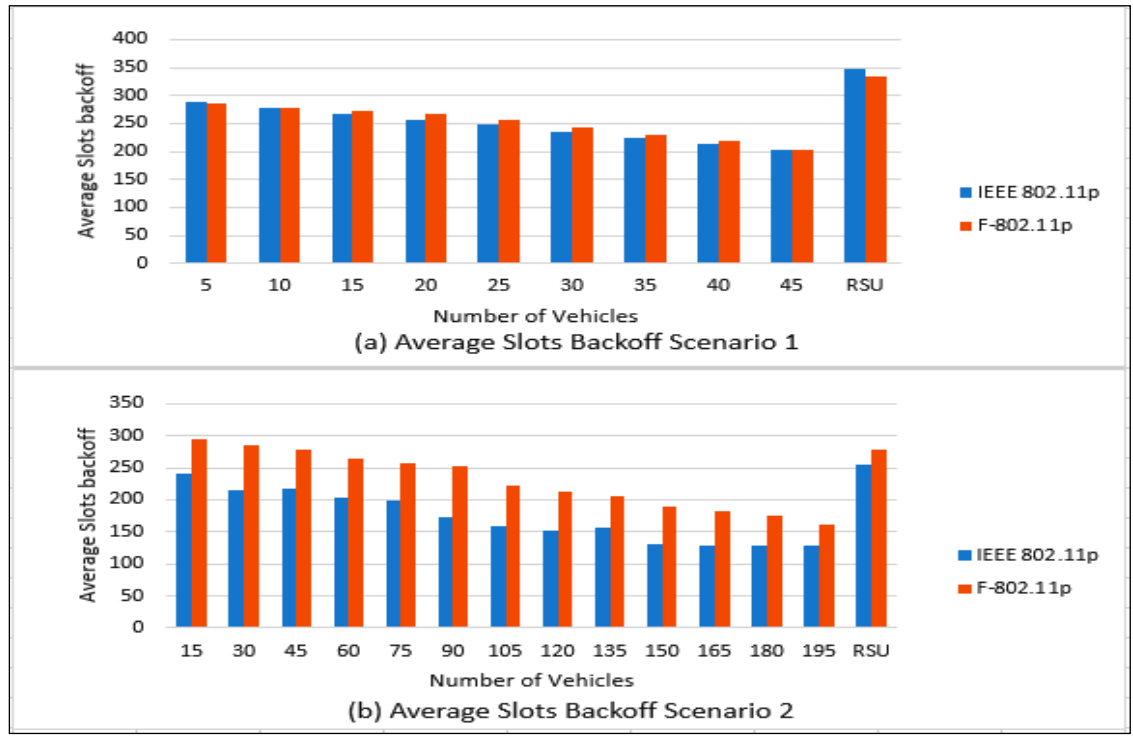

Figure 17. Slots backoff. (a) Scenario 1, (b) Scenario 2

## 6. CONCLUSION

In this paper, we presented F-802.11p, a fuzzy model that controls the broadcast of the WSM, WSA, and BSM messages. Before broadcasting these messages, the vehicle checks its status according to the F-802.11p. If the vehicle's status is within the F-802.11p allowable threshold, it processes the message, drops it, and waits. Consequently, the number of processed messages is decreased.

We have evaluated our proposed fuzzy mode against the IEEE 802.11p using OMNeT++, SUMO, Veins 5.0 framework for V2X, and MATLAB R2016b fuzzy toolbox in two scenarios. The first scenario uses the Veins framework. The second scenario uses the OpenStreetMap simulator. We conducted twelve experiments that measured the busy time of MAC and Physical layers, generated WSMs, BSMs, and WSAs sent packets, received WSM, BSM, and WSA, total lost packets, and time into backoff.

F-802.11p forms a scheme of arrangement between vehicles while sending/receiving airframes messages. This arrangement reduces the collided packets by 98%, redundant packets sent by 18.5%, network overhead by 98%, and increases channel idle time by 95% compared to IEEE 802.11p. In addition, F-802.11p improves the busy time, total busy without loss of safety, lost packets, and the number of backoff times. In general, F-802.11p is more suitable when vehicles' speed is restricted to 80 km/h.

During the simulation and evaluation of F-802.11p, intensive computational resources were required, synchronizing OMNET++ and SUMO was needed, and a lack of implementation steps in publications worked against numerical comparison. In the future, more factors and metrics may be considered in the fuzzy model to enhance the slots backoff and times into backoff decision-making.

**CONFLICT OF INTEREST**

The authors declare no conflict of interest.